\documentclass[12pt]{article}
\usepackage{theorem,amsfonts}
{\theoremstyle{break}
\theorembodyfont{\normalfont}
\newtheorem{theorem}{Theorem}[section]
\newtheorem{proposition}{Proposition}[section]}
\makeatletter
\@addtoreset{equation}{section}
\@addtoreset{table}{section}
\@addtoreset{figure}{section}
\makeatother

\renewcommand{\theequation}{\thesection.\arabic{equation}}

\def\Re{{\mathbb R}}
\def\Ce{{\mathbb C}}

\def\>#1{\mathbf{#1}}

\def\j{J}
\def\m{M}
\def\b{B}
\def\e{I}
\def\g{G}

\def\s{{\cal S}}
\def\G{{\cal G}}

\def\omm{X}

\def\jj{j}
\def\mm{m}
\def\bb{b}
\def\jm{jm}
\def\mj{mj}
\def\jb{jb}
\def\mb{mb}
\def\jjmm{jm}
\def\me{mi}
\def\je{ji}
\def\bee{bi}

\def\cext{\xi}

\def\cexta{\alpha}
\def\cextb{\beta}
\def\cextg{\gamma}
\def\cextj{\tau}
\def\cextm{\eta}

\def\bounj{\sigma}
\def\bounm{\rho}
\def\bounb{\upsilon}
\def\boune{\varsigma}

\def\be{\begin{equation}}
\def\ee{\end{equation}}
\def\bea{\begin{eqnarray}}
\def\eea{\end{eqnarray}}

\def\k{\omega}
\def\kbold{\omega}

\def\Lam{{{\cal I}_\kbold}}

\begin{document}

\thispagestyle{empty}

\hfill December 21, 1997
\bigskip\bigskip

\begin{center}

{\LARGE{\bf{Central Extensions of the families of
\\[0.3cm]
Quasi-unitary Lie algebras}}}
\end{center}

\bigskip

\begin{center}
F.J. Herranz$^\dagger$,
J.C. P\'erez Bueno$^\ddagger$
and M. Santander$^\star$
\end{center}

\begin{center}
{\it $^\dagger$ Departamento de F\'{\i}sica, E.U. Polit\'ecnica \\
Universidad de Burgos, E--09006 Burgos, Spain}
\end{center}

\begin{center}
{\it $^\ddagger$ Departamento de F\'{\i}sica Te\'orica and IFIC \\
Centro Mixto Universidad de Valencia--CSIC \\
E--46100 Burjassot, Valencia, Spain}
\end{center}

\begin{center}
{\it $^{\star}$ Departamento de F\'{\i}sica Te\'orica,
Universidad de Valladolid \\
E--47011, Valladolid, Spain}
\end{center}

\begin{abstract}
The most general possible central extensions of two whole families of  Lie
algebras, which can be obtained by contracting the special pseudo-unitary
algebras $su({p,q})$ of the  Cartan series $A_l$ and the pseudo-unitary
algebras
$u({p,q})$, are completely determined and classified for arbitrary $p,\ q$.
In addition to the $su({p,q})$ and $u({p,q})$ algebras, whose second cohomology
group is well known to be trivial, each family includes many non-semisimple
algebras; their central extensions, which are explicitly given, can be
classified into three types as far as their properties under contraction are
involved. A  closed expression for the dimension of the second cohomology group
of any member of these families of algebras is given.
\end{abstract}


\section{Introduction}

This paper investigates the Lie algebra cohomology of the unitary
Cayley--Klein (CK) families of Lie algebras
in any dimension.  These families, also called `quasi-unitary' algebras,
include both the special \mbox{(pseudo-)}unitary
$su({p,q})$ and \mbox{(pseudo-)}uni\-tary $u({p,q})$ algebras
---which have only trivial central
extensions\mbox{---,} as well as many other obtained from these  by a sequence
of contractions, which are no longer semisimple and may have
non-trivial central extensions.

The paper can be considered as a further step in a
series of studies on the CK families of Lie algebras. These
have both
mathematical interest and physical relevance. The families of CK
algebras provide a frame to describe the behaviour of mathematical
properties of algebras under contraction; in physical terms this is related to
some kind of approximation. The central extensions for the family of
quasi-orthogonal  algebras, also in the general situation and for any
dimension, have been determined in a previous paper \cite{Azc.Her.Bue.San:96}.
We refer to this work for references and for physical
motivations; we simply remark here that there are three main reasons behind the
interest in the second cohomology groups for Lie algebras. First, in any
quantum theory the relevant representations of any symmetry group are
projective instead of linear ones. Second, homogeneous symplectic manifolds
under a group appear as orbits of the coadjoint representation of either the
group itself or of a central extension. And third, quasi-invariant Lagrangians
are also directly linked to the central extensions of the group; these can be
related also to Wess--Zumino terms. In
addition to the references in
\cite{Azc.Her.Bue.San:96}, we may add that Wess--Zumino--Witten models leading
to central extensions have also been studied
(see \emph{e.g.}
\cite{Azc.Izq.Mac:90,Fig.Sta:94} and references therein).

The knowledge of the second cohomology
group for a Lie algebra relies on the general solution of a set of linear
equations, yet some general results allow to bypass the calculations in
special cases. For instance, the second cohomology group is trivial for
semisimple Lie algebras. But once a contraction is made, the semisimple
character disappears, and the contracted algebra might have non-trivial central
extensions. Instead of finding the general solution for the extension equations
on a case-by-case basis, our approach is to do these calculations for a
whole family including a large number of algebras simultaneously. This program
has been developed for the quasi-orthogonal algebras, and here we discuss the
`next' quasi-unitary case. There are two main advantages in this approach.
First, it allows to record, in a form easily retrievable, a large number of
results which can be needed in applications, both in mathematics and in
physics. This avoids at once and for all the case-by-case type computation of
the
central extensions of algebras included in the unitary families. And second, it
sheds some further light on the interrelations between cohomology and
contractions, by discussing in particular examples how and when a contraction
increases the cohomology of the algebra: central extensions can be classed into
three types, with different behaviour under contraction.

The section~\ref{sec.2} is devoted to the description of the  two families of
unitary CK algebras. We show how to obtain these as graded contractions of the
compact algebras $su(N+1)$ and
$u(N+1)$, and we provide some details on their structure. It should be remarked
that the CK unitary algebras are associated to the complex hermitian spaces
with metrics of different signatures and to their contractions. In
section~\ref{sec.3} the general solution to the central extension problem for
these algebras is given; this includes the completely explicit description of
all possible central extensions and the discussion of their triviality. A
closed formula for the dimension of the second cohomology group is also
obtained. Computational details on the procedure to
solve the central extension problem are given in an Appendix. The results are
illustrated in section~\ref{sec.4} for the lowest dimensional examples.
Finally, some remarks close the paper.


\section{The CK families of quasi-unitary algebras}
\label{sec.2}

The family of special quasi-unitary algebras, which involves the simple
Lie algebras $su(p,q)$, as well as many non-simple algebras obtained
by \.In\"on\"u--Wigner \cite{Ino.Wig:53} contraction from $su(p,q)$ can be
easily described in terms of graded contraction theory
\cite{Mon.Pat:91,Moo.Pat:91}, taking the compact real form
$su(N+1)$ of the simple algebras in the series
$A_N$ as starting point. As it is well known, the special unitary algebra
can be realised by complex antihermitian and traceless matrices, and is the
quotient of the algebra of all complex antihermitian matrices by its
center (generated by the pure imaginary multiples of the identity). It
 will be convenient to consider the family of quasi-unitary algebras
altogether; these can be similarly described in terms of graded
contractions of $u(N+1)$, and will include algebras obtained from $u(p,q)$
by \.In\"on\"u--Wigner contractions. Let us consider the (fundamental)
matrix representation of the algebras
$su(N+1)$ and
$u(N+1)$, as given by the complex matrices
$\j_{ab},\ \m_{ab},\ \b_l$ and $\j_{ab},\ \m_{ab},\ \b_l,\ \e$:
\be
\j_{ab}=-e_{ab}+e_{ba}  \quad
\m_{ab}=i(e_{ab}+e_{ba})  \quad
\b_l=i(e_{l-1,l-1}-e_{ll})  \quad
\e=i \sum_{a=0}^{N}e_{aa} ,
\label{usuRealMat}
\ee
where $a<b$, $a,b=0,\dots,N$,\  $l=1,\dots,N$, and where $e_{ab}$ means
the $(N+1) \times (N+1)$ matrix with a single  1 entry in row $a$ and
column $b$. The commutation relations involved in either of these
algebras are given by
\be
\begin{array}{lll}
[\j_{ab},\j_{ac}] = \j_{bc} &\qquad
[\j_{ab},\j_{bc}] =-\j_{ac} &\qquad
[\j_{ac},\j_{bc}] = \j_{ab}\\{}
[\m_{ab},\m_{ac}] = \j_{bc} &\qquad
[\m_{ab},\m_{bc}] = \j_{ac} &\qquad
[\m_{ac},\m_{bc}] = \j_{ab} \\{}
[\j_{ab},\m_{ac}] = \m_{bc} &\qquad
[\j_{ab},\m_{bc}] =-\m_{ac} &\qquad
[\j_{ac},\m_{bc}] =-\m_{ab}\\{}
[\m_{ab},\j_{ac}] =-\m_{bc} &\qquad
[\m_{ab},\j_{bc}] =-\m_{ac} &\qquad
[\m_{ac},\j_{bc}] = \m_{ab} \\{}
[\j_{ab},\j_{de}]= 0 &\qquad
[\m_{ab},\m_{de}] =0 &\qquad
[\j_{ab},\m_{de}] =0 \\
\multicolumn{3}{c}{
  [\j_{ab},\b_l] = ( \delta_{a,l-1} -\delta_{b,l-1}  +
   \delta_{bl} -\delta_{al})\m_{ab}} \\
\multicolumn{3}{c}{
  [\m_{ab},\b_l] = - ( \delta_{a,l-1} -\delta_{b,l-1}  +
   \delta_{bl} -\delta_{al})\j_{ab}}
\end{array}
\label{usuCR}
\ee
\be
[\j_{ab},\m_{ab}] =-2\sum_{s=a+1}^b \b_s  \qquad\qquad
[\b_{k},\b_{l}]=0
\label{suCR}
\ee
\be
\begin{array}{lll}
[\j_{ab},\e] = 0 & \qquad
[\m_{ab},\e] = 0 & \qquad
[\b_l,\e] = 0 .
\end{array}
\label{uCR}
\ee

The algebra $su(N+1)$ has a grading by a group
$\mathbb{Z}_2^{\otimes N}$
related to a set of $N$ commuting involutions in the
subalgebra $so(N+1)$ generated by $\j_{ab}$
\cite{Her.Mon.Olm.San:94,Her.San:96b}.
If $\s$ denotes any subset of the set of indices $\{0, 1, \dots, N\}$,
and $\chi_\s(a)$ denotes the characteristic function over $\s$,
then each of the linear mappings given by
\be
S_\s \j_{ab} = (-1)^{\chi_\s(a) + \chi_\s(b)} \j_{ab}  \qquad
S_\s \m_{ab} = (-1)^{\chi_\s(a) + \chi_\s(b)} \m_{ab}  \qquad
S_\s \b_{l} =  \b_{l}
\ee
is an involutive automorphism of the algebra $su(N+1)$; by
considering all possible subsets of indices we get $2^N$ different
automorphisms defining a $\mathbb{Z}_2^{\otimes N}$ grading for this
algebra. The corresponding graded contractions of
$su(N+1)$ constitute a large set of Lie algebras, but there exists
a particular subset or family of these graded contractions, nearer
to the simple ones, which essentially
preserves the properties
associated to simplicity, and which belong to the so termed
\cite{Roz:88, Roz:97} `quasi-simple' algebras. This family, to be defined
below, encompasses the special pseudo-unitary algebras (in the $A_N$ Cartan
series)  as well as their nearest non-simple contractions. By taking the
generator $\e$ as invariant under all involutions, this grading can
be extended to the algebra $u(N+1)$, whose graded contractions include the
pseudo-unitary algebras as well as many non-semisimple algebras;  again a
particular family of these graded contractions, to be introduced below,
preserves properties associated to semi-simplicity.  Collectively, all these
algebras (special or not) are called
\emph{quasi-unitary}; these are also called Cayley--Klein algebras of unitary
type, or unitary CK algebras, since they are exactly those
algebras behind the geometries of a complex hermitian space  with a
projective metric in the CK sense \cite{Roz:97}. Another
view to these algebras is given in \cite{Gro.Man:90}.

The overall details on the structure of this family are similar
to the orthogonal case. The set of unitary CK algebras is parametrised by
$N$ real coefficients $\k_a$ ($a=1,\dots,N$), whose values codify in a
convenient way the pertinent information on the Lie algebra structure
\cite{Her:95,Her.San:96}. In terms of the $N(N+1)/2$
two-index coefficients
$\k_{ab}$ defined by
\be
\k_{ab}:=\k_{a+1}\k_{a+2}\cdots\k_b  \qquad
  a,b=0,1,\dots,N  \quad
  a<b  \qquad
\k_{aa}:=1
\label{aa}
\ee
which verify
\be
\k_{ac}=\k_{ab}\k_{bc}  \qquad
  a \leq b \leq c  \qquad
\k_{a}=\k_{a-1\, a}  \qquad
  a=1,\dots,N,
\label{ab}
\ee
the algebras to be denoted $su_{\kbold}(N+1)$ and $u_{\kbold}(N+1)$,
$\kbold \equiv (\k_1,\dots,\k_N)$, of
dimensions $(N+1)^2-1$ and $(N+1)^2$, are generated by $\j_{ab},\
\m_{ab},\ \b_l$ and $\j_{ab},\ \m_{ab},\ \b_l,\ \e$ ($a<b$), with commutators:
\be
\begin{array}{lll}
[\j_{ab},\j_{ac}] =\k_{ab}\j_{bc} &\qquad
[\j_{ab},\j_{bc}] =-\j_{ac} &\qquad
[\j_{ac},\j_{bc}] =\k_{bc}\j_{ab}\\{}
[\m_{ab},\m_{ac}] =\k_{ab}\j_{bc} &\qquad
[\m_{ab},\m_{bc}] =\j_{ac} &\qquad
[\m_{ac},\m_{bc}] =\k_{bc}\j_{ab} \\{}
[\j_{ab},\m_{ac}] =\k_{ab}\m_{bc} &\qquad
[\j_{ab},\m_{bc}] =-\m_{ac} &\qquad
[\j_{ac},\m_{bc}] =-\k_{bc}\m_{ab}\\{}
[\m_{ab},\j_{ac}] =-\k_{ab}\m_{bc} &\qquad
[\m_{ab},\j_{bc}] =-\m_{ac} &\qquad
[\m_{ac},\j_{bc}] =\k_{bc}\m_{ab} \\{}
[\j_{ab},\j_{de}]= 0 &\qquad
[\m_{ab},\m_{de}] =0 &\qquad
[\j_{ab},\m_{de}] =0 \\
\multicolumn{3}{c}{
   [\j_{ab},\b_l] = ( \delta_{a,l-1} -\delta_{b,l-1}  +
    \delta_{bl} -\delta_{al})\m_{ab}} \\
\multicolumn{3}{c}{
   [\m_{ab},\b_l] = - ( \delta_{a,l-1} -\delta_{b,l-1}  +
    \delta_{bl} -\delta_{al})\j_{ab}} \\
\end{array}
\label{CKusuCR}
\ee
\be
[\j_{ab},\m_{ab}] =-2\k_{ab}\sum_{s=a+1}^b \b_s  \qquad\qquad
[\b_{k},\b_{l}]=0
\label{CKsuCR}
\ee
\be
\begin{array}{lll}
[\j_{ab},\e] = 0 & \qquad
[\m_{ab},\e] = 0 & \qquad
[\b_l,\e] = 0
\label{CKuCR}
\end{array}
\ee
where $a, b, c, d, e=0, \dots, N$ and $k, l=1, \dots, N$; we assume
$a<b<c$ for each set of three  indices  $\{a,b,c\}$, and $a<b,\ d<e$
for each set of four indices $\{a,b,d,e\}$ which are also assumed to
be \emph{different}.

\subsection{The unitary CK groups}

The connection with groups of isometries of a hermitian metric is as
follows: for a generic choice, with \emph{all} $\k_a \neq 0$, let us
consider the space $\Ce^{N+1}$ endowed with a hermitian
(sesqui)linear form $\langle .|. \rangle_\kbold : \Ce^{N+1}\times
\Ce^{N+1}\to  \Ce$ associated to the matrix
\be
\Lam = {\mbox{diag}}\ (1,\, \k_{01},\,\k_{02},\dots,\,\k_{0N}) =
       {\mbox{diag}}\ (1,\, \k_1,\,\k_1\k_2,\dots,\,\k_1\cdots\k_N);
\label{suMetricMatrix}
\ee
this is, for any pair of vectors $\>a,\>b\in \Ce^{N+1}$,
\be
\langle \>a|\>b \rangle_\kbold:=
\bar a^0 b^0 + \bar a^1 \k_1 b^1 + \bar a^2 \k_1 \k_2 b^2 + \dots =
\sum_{i=0}^N \bar a^i\k_{0i}  b^i.
\label{ai}
\ee
Let us define the group $U_{\k_1,\dots,\k_N}(N+1)\equiv
U_{\kbold}(N+1)$ as the group of linear isometries of
the hermitian metric (\ref{suMetricMatrix}). The isometry
condition
\be
\langle U \>a| U \>b \rangle_\kbold= \langle \>a|  \>b \rangle_\kbold
\qquad \forall\, \>a,\>b\in \Ce^{N+1},
\label{aj}
\ee
implies for the matrix
$U \in U_{\kbold}(N+1)$ the condition
\be
U^\dagger\Lam U=\Lam  \qquad
\forall U\in U_{\kbold}(N+1).
\label{ak}
\ee
For the corresponding Lie algebra the above relation leads to
\be
X^\dagger\Lam +\Lam X=0  \qquad
\forall X\in  u_{\kbold}(N+1).
\label{ag}
\ee
This Lie algebra is generated by
the complex matrices (cf. (\ref{usuRealMat}))
\be
\j_{ab}=-\k_{ab}e_{ab}+e_{ba}  \quad
\m_{ab}=i(\k_{ab}e_{ab}+e_{ba})  \quad
\b_l=i(e_{l-1,l-1}-e_{ll})  \quad
\e=i \sum_{a=0}^{N}e_{aa}
\label{CKusuRealMat}
\ee
with $a<b$, $a,b=0,\dots,N$, \ $l=1,\dots,N$.

The group $SU_{\k_1,\dots,\k_N}(N+1)\equiv SU_{\kbold}(N+1)$ is
defined similarly by adding the unimodularity condition $\det(U)=1$;
this leads for the Lie algebra to the condition ${\mbox{trace}}(X)
=0$, so the algebra $su_{\kbold}(N+1)$ is generated by
$\j_{ab}, \m_{ab}, \b_l$ alone.

The action of the groups $U_{\kbold}(N+1)$ and $SU_{\kbold}(N+1)$ in
$\Ce^{N+1}$ is not transitive, and the `sphere' with equation
\be
\langle \>x|\>x \rangle_\kbold:= \sum_{i=0}^N \bar x^i\k_{0i}  x^i =1
\label{suSphere}
\ee
is stable. For the action of
$SU_{\kbold}(N+1)$, the isotropy subgroup of a reference point in this
sphere, say $(1, 0, \dots, 0)$, is easily shown to be isomorphic to
$SU_{\k_2,\k_3, \dots, \k_N}(N)$, and the isotropy subgroup of the
\emph{ray} of a reference point is
$U_{\k_2,\k_3, \dots, \k_N}(N)$, locally isomorphic to $U(1) \otimes
SU_{\k_2,\k_3,
\dots, \k_N}(N)$. The quotient spaces
$SU_{\k_1, \k_2, \k_3, \dots, \k_N}(N+1)/\big( U(1) \otimes SU_{\k_2,\k_3,
\dots, \k_N}(N) \big) $ are a family of  hermitian spaces which
includes examples with non-definite and/or
degenerate hermitian metrics; the CK scheme provides a common frame to discuss
all them jointly. The most familiar corresponds to $\k_2=\k_3=\dots=\k_N=1$,
and depends on a single parameter $\k_1=K$; when $K>0$ or $K<0$ these
are the usual elliptic or hyperbolic complex hermitian spaces of (holomorphic
constant) curvature $K$; when $\k_1=0$ we get
the `Euclidean' flat hermitian space (finite-dimensional Hilbert space).

When the constants $\k_a$ are allowed to vanish, the set of
isometries of the hermitian metric (\ref{suMetricMatrix}) is larger
than the group generated by the matrices
$\j_{ab},\ \m_{ab},\ \b_l,\ \e$. In this case, there are additional
geometric structures in $\Ce^{N+1}$ (related to the existence of
additional invariant foliations similar to the one implied by
(\ref{suSphere})), and the proper definition of the automorphism
group of these structures leads again to the group generated by the
matrix Lie algebra (\ref{CKusuRealMat}) with the commutation relations
 (\ref{CKusuCR})--(\ref{CKuCR}). These matrix
realisations can be considered as the fundamental representation of
the  unitary CK Lie algebras
$su_{\kbold}(N+1)$ and $u_{\kbold}(N+1)$.

Since each coefficient $\k_a$ can be positive, negative or zero,
each unitary CK family comprises  $3^N$ Lie algebras  although some of
them may be isomorphic. For instance, the map
\be
\j_{ab}\to \j'_{ab}=-\j_{N-b,N-a}  \quad
\m_{ab}\to \m'_{ab}=-\m_{N-b,N-a}  \quad
\b_{l}\to \b'_{l}= \b_{N+1-l}
\label{aga}
\ee
provides an isomorphism
\be
su_{\k_1,\k_2,\dots,\k_{N-1},\k_N}(N+1)\simeq
su_{\k_N,\k_{N-1},\dots,\k_2,\k_1}(N+1)  .
\label{PolarityIsom}
\ee

\subsection{Structure of the unitary CK algebras}

The  unitary CK algebras $su_{\kbold}(N+1)$ contain many subalgebras
isomorphic to algebras in both families $su_{\kbold}(M+1)$ and
$u_{\kbold}(M+1)$, $M<N$. To best describe this,  we introduce a new set
of Cartan subalgebra generators for $su_{\kbold}(N+1)$,
$\g_a$ ($a=1,\dots,N$), defined by
\be
\begin{array}{rl}
\g_a:= & \displaystyle
\frac{1}{a} \Big( \b_1+2\b_2+\dots+(a-1)\b_{a-1}\Big) + B_a
\\
+ &
\displaystyle
\frac{1}{N+1-a} \Big( (N-a)\b_{a+1}+(N-a-1)\b_{a+2}+\dots+\b_{N}\Big).
\end{array}
\label{Ggenerators}
\ee
In the matrix realisation (\ref{CKusuRealMat}) $\g_a$ is given by
\be
\g_a=i\Big(\frac{1}{a}(\sum_{s=0}^{a-1}e_{ss})-
           \frac{1}{N+1-a}(\sum_{s=a}^{N}e_{ss})\Big),
\ee
so each $\g_a$ appears as a direct sum of two blocks,
each proportional with a pure imaginary coefficient to the identity matrix.

Denoting by $\omm_{ij}$ the pair of generators
$\{\j_{ij},\m_{ij}\}$, we can check that the set $\langle
\omm_{ij},\ i,j=0,1,\dots,a-1\ ;\ \b_l,\ l=1,\dots,a-1\rangle$
closes a Lie subalgebra
$su_{\k_1,\dots,\k_{a-1}}(a)$.
Furthermore,
$\g_a$ commutes with all the generators in this subalgebra, so that
the former generators plus $a\g_a$ close an algebra isomorphic to
$u_{\k_1,\dots,\k_{a-1}}(a)$.

Similarly, the set $\langle \omm_{ij},\ i,j=a,a+1,\dots,N;\ \b_l,\
l=a+1,\dots,N \rangle$ closes the  special unitary CK Lie algebra
$su_{\k_{a+1},\dots,\k_N}(N+1-a)$, and by adding $-(N+1-a)\g_a$ we
get an algebra isomorphic to $u_{\k_{a+1},\dots,\k_N}(N+1-a)$.

\begin{figure}[ht]
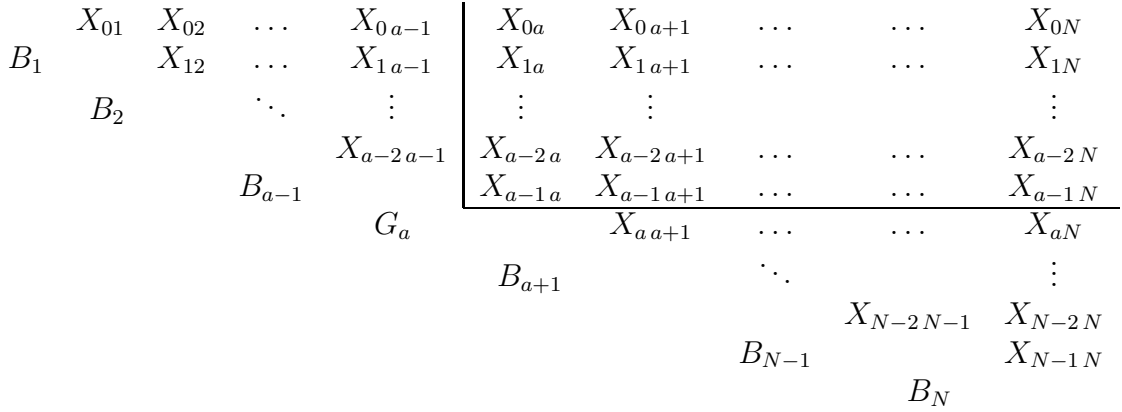

\begin{center}
\begin{tabular}{ccccc|cccccc}
& $\omm_{01} $&$ \omm_{02} $&$\ldots$&$\omm_{0\, a-1} $&
   $\omm_{0a}$&$\omm_{0\, a+1}$&$\ldots$&$\ldots$&$\omm_{0N}$\\
$\hfill \b_{1} $ && $ \omm_{12} $&$\ldots$&$\omm_{1\, a-1} $&
   $\omm_{1a}$&$\omm_{1\, a+1}$& $\ldots$&$\ldots$&$\omm_{1N}$\\
& $\hfill \b_{2} $ &$  $&$\ddots $&$\vdots$&
   $\vdots$&$\vdots$&$ $&&$\vdots$\\
& &$ $&$ $&$\omm_{a-2\,a-1}$&  $\omm_{a-2\,a}$&$\omm_{a-2\,a+1}$&
  $\ldots$&$\ldots$&$\omm_{a-2\,N}$\\
& &$ $& $\hfill \b_{a-1}$ &&  $\omm_{a-1\,a}$&$\omm_{a-1\,a+1}$&
   $\ldots$&$\ldots$&$\omm_{a-1\,N}$\\
\cline{6-10}
& &$ $&\multicolumn{2}{c}{\,} $\hfill \g_{a}\ \ \ \ $
   &&$\omm_{a\,a+1}$& $\ldots$&$\ldots$&$\omm_{a  N}$\\
& &$ $&\multicolumn{2}{c}{\,}&  $\hfill \b_{a+1}$&
  $ $& $\ddots$&&$\vdots$\\
& &$ $&\multicolumn{2}{c}{\,}&  $ $&
  $ $ $  $&&$\omm_{N-2\,N-1}$&$\omm_{N-2\,N}$\\
& &$ $&\multicolumn{2}{c}{\,}&  $ $&
  $ $ &$\b_{N-1}$&&$\omm_{N-1\,N}$\\
& &$ $&\multicolumn{2}{c}{\,}&  $ $& $ $ &&$\hfill \b_{N}\ \ $\\
\end{tabular}
\end{center}
\caption{Generators of the (special) unitary CK algebras}
\label{fig2.1}
\end{figure}

This structure can be visualised by arranging the basis generators
as in Fig.~\ref{fig2.1}.

The special unitary subalgebras $su_{\k_1,\dots,\k_{a-1}}(a)$  and
$su_{\k_{a+1},\dots,\k_N}(N+1-a)$ correspond, in this order, to the
two triangles to the left and below the rectangle, both excluding the
generator $\g_a$.  The unitary subalgebras
$u_{\k_1,\dots,\k_{a-1}}(a)$  and $u_{\k_{a+1},\dots,\k_N}(N+1-a)$
correspond, in this order, to the two triangles to the left and
below the rectangle, both including the generator $\g_a$.
This generator $\g_a$ closes a $u(1)$ subalgebra.

We sum up the details relative to the structure of the special unitary
CK algebras in two statements
\begin{itemize}
\item
When all $\k_a$  are different from zero,
$su_{\kbold}(N+1)$ is a  pseudo-unitary simple Lie algebra
$su({p,q})$ in the  Cartan series $A_N$ ($p$ and $q$ are the number
of positive and negative signs in diagonal of the metric matrix
(\ref{suMetricMatrix}), $p+q=N+1$).

\item
If a coefficient $\k_a$ vanishes, the CK algebra is a
non-simple Lie algebra which has a semidirect structure
\be
\begin{array}{l}
su_{\k_1,\dots,\k_{a-1},\k_a=0,\k_{a+1},\dots,\k_N}(N+1)
\equiv \\
\qquad\qquad
t \odot
( su_{\k_1,\dots,\k_{a-1}}(a)
\oplus
u(1)
\oplus
su_{\k_{a+1},\dots,\k_N}(N+1-a)),
\end{array}
\label{IsotropyDecom}
\ee
where the subalgebras appearing in (\ref{IsotropyDecom}) are
generated by
\be
\begin{array}{l}
t=\langle \omm_{ij},\ i=0,1,\dots,a-1,\
             j=a,a+1,\dots,N\rangle \cr
su_{\k_1,\dots,\k_{a-1}}(a) =\langle
    \omm_{ij},\ i,j=0,1,\dots,a-1\ ;\
    \b_l,\ l=1, \dots ,a-1 \rangle \cr
u(1) = \langle \g_a \rangle \cr
su_{\k_{a+1},\dots,\k_N}(N+1-a) =\langle
    \omm_{ij},\, i,j=a,a+1,\dots,N;\,
    \b_l,\ l=a+1, \dots,N \rangle.
\end{array}
\label{aia}
\ee
We note that $t$ is an abelian subalgebra of dimension $2a(N+1-a)$. In
terms of the triangular arrangement of generators (Fig.~\ref{fig2.1}), $t$
is spanned by the generators inside the rectangle; we remark that these
generators do not close a subalgebra when $\k_a\neq 0$. The three remaining
sets are always subalgebras, no matter of whether or not $\k_a=0$.
\end{itemize}

For the particular case $\k_1=0$ (or, \emph{mutatis mutandis}, $\k_N=0$) the
contracted algebra is a quasi-unitary inhomogeneous algebra,
$$
su_{0,\k_2,\dots,\k_N}(N+1)
\equiv t_{2N}\odot  u_{\k_2,\dots,\k_N}(N).
$$
The subindex $2N$ in $t$ denotes the real dimension of
$t \equiv \Ce^N$ which can be identified with the space
$SU_{0, \k_2, \k_3, \dots, \k_N}(N+1) / U_{\k_2,\dots,\k_N}(N)$, with the
natural action of  $U_{\k_2,\dots,\k_N}(N)$ (locally isomorphic to $
U(1) \otimes SU_{\k_2, \k_3, \dots, \k_N}(N)$) over $\Ce^N$. This
direct product appeared as the isotropy subalgebra of a ray for the natural
action of $SU_{0, \k_2, \k_3, \dots, \k_N}(N+1)$ on $\Ce^{N+1}$ discussed after
(\ref{suSphere}). In the case where
$\k_2, \k_3, \dots, \k_N$ are all different from zero, the algebra is an
ordinary inhomogeneous pseudo-unitary (not special) algebra:
$$
t_{2N}\odot  u_{\k_2,\dots,\k_N}(N) \equiv iu({p,q})
\qquad p+q=N \
$$
and in this case
$t_{2N}$ can be identified to the
$N$-dimensional flat complex hermitian space with signature
$p,q$ determined as the number of positive and negative terms in the sequence
$(1, \k_2, \k_2\k_3, \dots, \k_2\dots\k_N)$.

When several coefficients $\k_a$ are equal to zero the algebra
$su_{\k_1,\k_2,\dots,\k_N}(N+1)$ has simultaneously several such
decompositions. The more contracted case corresponds
to taking all $\k_a$ equal to zero; this gives rise to the  special
unitary flag algebra.


\section{Central extensions}
\label{sec.3}

Now we proceed to compute in a unified way all the central extensions
for the two unitary families of CK algebras, for arbitrary choices of
the constants $\k_a$ and  in any dimension. Let $\G$  be an arbitrary
$r$-dimensional Lie algebra with generators $\{X_1,\dots,X_r\}$ and
structure constants
$C_{ij}^k$. A central extension   $\overline{\G}$ of the algebra $\G$
by the one-dimensional algebra generated by
 $\Xi$ will have $(r+1)$ generators
$( X_i,\Xi) $ with commutation relations given by
\be
[X_i,X_j]=\sum_{k=1}^r C_{ij}^k X_k  + \cext_{ij} \Xi  \qquad
[\Xi,X_i]=0 .
\label{CentExt}
\ee
The   extension coefficients or  central charges
$\cext_{ij}$  must be antisymmetric in the indices $i,j$,
$\cext_{ji}=-\cext_{ij}$ and must fulfil the following conditions
coming from the Jacobi identities in the extended Lie algebra:
\be
\sum_{k=1}^r
\left(
C_{ij}^{k}\cext_{kl}+C_{jl}^{k}\cext_{ki}+C_{li}^{k}\cext_{kj}
\right) =0 .
\label{JacobiEqs}
\ee

These  extension coefficients  are the coordinates
$(\cext(X_i,X_j)=\cext_{ij})$ of the antisymmetric
two-tensor $\cext$ which is the two-cocycle of the specific extension
being considered, and  (\ref{JacobiEqs}) is
the two-cocycle condition for the Lie algebra cohomology.

Let us consider  the `abstract' extended Lie algebra $\overline{\G}$
with the  Lie brackets (\ref{CentExt}) and let us perform a change of
generators:
\be
X_i\to X'_i=X_i+\mu_i\Xi,
\label{ChangeGens}
\ee
where $\mu_i$ are arbitrary real
numbers. The commutation rules
for the generators $\{X'_i\}$ become
\be
[X'_i,X'_j]=\sum_{k=1}^r
C_{ij}^k X'_k  + (\cext_{ij} -\sum_{k=1}^r C_{ij}^k \mu_k )\Xi .
\label{ChangeGensCR}
\ee
Thus, the general expression for the two-coboundary $\delta\mu$
generated  by $\mu$ is
\be
\delta\mu(X_i, X_j) = \sum_{k=1}^r C_{ij}^k
\mu_{k}  .
\label{Cobound}
\ee
Two two-cocycles differing by a two-coboundary lead to equivalent
extensions; the classes of equivalence of non-trivial two-cocycles
associated with the tensors $\cext$ determine the second cohomology group
$H^2(\G,\Re)$.

\subsection{The  general solution to the extension problem for the
 unitary CK algebras}

In a previous paper \cite{Azc.Her.Bue.San:96}
we have given the general solution to
the extension equations for the case of the  orthogonal CK algebras.
The same approach can be used for the   family of
quasi-unitary algebras. However, and in order not to burden the
exposition, the main details on the procedure have been placed in the Appendix.
The results obtained there give the  general solution to the problem of
finding the central extensions for the unitary CK
algebras. They are summed up as

\begin{theorem}
\label{theor3.1}
The most general central extension $\overline{su}_\kbold(N+1)$ of
any algebra in the family of special unitary CK algebras
${su}_\kbold(N+1)$ is determined by the following \emph{basic}
coefficients:
\begin{description}
\item[Type I.]
$N(N+1)/2$ basic extension coefficients $\cextm_{ab}$ and
$N(N+1)/2$ basic extension coefficients $\cextj_{ab}$  ($a<b$,
$a,b=0,1,\dots,N$). These coefficients are not subjected to any
further relationship.

\item[Type II.] $N$ basic extension coefficients $\cexta_{k}$
($k=1,\dots,N$), not subjected to any further relationship.

\item[Type III.] $N(N-1)/2$ basic extension coefficients $\cextb_{kl}$
($k<l$, $k,l=1,\dots,N$) which must satisfy the conditions
\be
\k_k \cextb_{kl}=0  \qquad \k_l\cextb_{kl}=0 .
\label{ECtaRestrict}
\ee
\end{description}
\end{theorem}

\begin{theorem}
\label{theor3.2}
The most general central extension $\overline{u}_\kbold(N+1)$ of
any algebra in the unitary CK family
${u}_\kbold(N+1)$, is determined by the basic extension
coefficients given in Theorem~\ref{theor3.1}, and by an additional set of

\begin{description}
\item[Type III.]
$N$ basic extension coefficients $\cextg_{k}$
($k=1,\dots,N$), subjected to the relation
\be
\k_k \cextg_{k}=0.
\label{urestrict}
\ee
\end{description}
\end{theorem}
For any given choice of the constants $\k_a$, these basic extension
coefficients determine two-cocycles for the algebras
${su}_\kbold(N+1)$ and
${u}_\kbold(N+1)$. The Lie brackets of the extended algebras
$\overline{su}_\kbold(N+1)$ and $\overline{u}_\kbold(N+1)$ are
given by
\be
\begin{array}{ll}
[\j_{ab},\j_{ac}] =\k_{ab}(\j_{bc}+\cextm_{bc}\Xi) &\qquad
[\m_{ab},\m_{ac}] =\k_{ab}(\j_{bc}+\cextm_{bc}\Xi)\cr
[\j_{ab},\j_{bc}] =-(\j_{ac} +\cextm_{ac}\Xi) &\qquad
[\m_{ab},\m_{bc}] =\j_{ac}+\cextm_{ac}\Xi\cr
[\j_{ac},\j_{bc}] =\k_{bc}(\j_{ab} + \cextm_{ab}\Xi) &\qquad
[\m_{ac},\m_{bc}] =\k_{bc}(\j_{ab} + \cextm_{ab}\Xi) \cr
[\j_{ab},\j_{mn}]=0 &\qquad
[\m_{ab},\m_{mn}] =0\cr
[\j_{ab},\m_{ac}] =\k_{ab}(\m_{bc}+\cextj_{bc} \Xi) &\qquad
[\m_{ab},\j_{ac}] =-\k_{ab}(\m_{bc} +  \cextj_{bc}\Xi)
\cr
[\j_{ab},\m_{bc}] =-(\m_{ac}+ \cextj_{ac} \Xi) &\qquad
[\m_{ab},\j_{bc}] =-(\m_{ac} + \cextj_{ac} \Xi)
\cr
[\j_{ac},\m_{bc}] =-\k_{bc}(\m_{ab} +\cextj_{ab}\Xi) &\qquad
[\m_{ac},\j_{bc}] =\k_{bc}(\m_{ab} + \cextj_{ab} \Xi)
\cr
[\j_{ab},\m_{mn}] =0 &\qquad
[\m_{ab},\j_{mn}] =0 \cr
\multicolumn{2}{l}{
[\j_{ab},\b_l] = ( \delta_{a,l-1} -\delta_{b,l-1}  +
\delta_{bl} -\delta_{al}) (\m_{ab} + \cextj_{ab} \Xi)}\cr
\multicolumn{2}{l}{
[\m_{ab},\b_l] = - ( \delta_{a,l-1} -\delta_{b,l-1}  +
\delta_{bl} -\delta_{al})(\j_{ab}   + \cextm_{ab} \Xi)}\cr
\multicolumn{2}{l}
{\displaystyle
[\j_{ab},\m_{ab}]=-2\k_{ab}\sum_{s=a+1}^b \b_s +
\sum_{s=a+1}^b\k_{a\,s-1}\k_{sb}\,\cexta_{s} \Xi  \qquad
[\b_k,\b_l]=\cextb_{kl}\Xi}
\end{array}
\label{CKusuGenExtCR}
\ee
\be
\begin{array}{lll}
[\j_{ab},\e] = 0  & \qquad
[\m_{ab},\e] = 0  & \qquad
[\b_k,\e] =\cextg_k \Xi,
\label{CKuGenExtCR}
\end{array}
\ee
where $a<b<c$, $k<l$, $m<n$  and $a,b,m,n$ are all different.

The complete expression for the two-cocycles for ${su}_\kbold(N+1)$ and
${u}_\kbold(N+1)$ can be read directly
from these commutators; for future convenience, we collect some
expressions relating the basic extension coefficients with
particular values of the two-cocycles determining the extensions
(however, and as it can be seen in (\ref{CKusuGenExtCR}), most of
these basic coefficients appears related to the values of the
cocycle in several ways)
\be
\begin{array}{ll}
\cextm_{ac} = -\cext(\j_{ab}, \j_{bc})  & \quad
  \cextj_{ac} = -\cext(\j_{ab},  \m_{bc}) \\
\cexta_{k} = \cext(\j_{k-1\,k}, \m_{k-1\, k}) & \quad
  \cextb_{kl} = \cext(\b_{k}, \b_{l})
\end{array}
\label{suBasicExtCoef}
\ee
\be
 \cextg_k = \cext(\b_k, \e).
\label{suBasicExtCoefb}
\ee

\subsection{Equivalence of extensions}

According to the general discussion in the beginning of this section,
we now look for the more general coboundary for ${su}_\kbold(N+1)$ or
${u}_\kbold(N+1)$. We write a change of basis (see (\ref{ChangeGens}))
for the generators as
\be
\j_{ab}\to \j'_{ab}=\j_{ab}+\bounj_{ab}\Xi  \quad
\m_{ab}\to \m'_{ab}=\m_{ab}+\bounm_{ab}\Xi  \quad
\b_{k}\to \b'_{k}=\b_{k}+\bounb_{k}\Xi
\label{ddequiv}
\ee
\be
\e \to \e + \boune \Xi
\ee
where $\bounj_{ab},\ \bounm_{ab},\ \bounb_{k},\ \boune$ are the values
of $\mu$ on the generators $\j_{ab},\ \m_{ab},\ \b_{k},\ \e$.
By using (\ref{Cobound}) and the structure constants of the algebras
${su}_\kbold(N+1)$ or ${u}_\kbold(N+1)$
read from (\ref{CKusuCR})--(\ref{CKuCR}),
we find for the associated coboundaries $\delta\mu$,
\be
\begin{array}{ll}
\delta\mu(\j_{ab}, \j_{bc}) = - \bounj_{ac} & \qquad
\delta\mu(\j_{ab}, \m_{bc}) = - \bounm_{ac} \\
\delta\mu(\j_{k-1\, k}, \m_{k-1\, k}) = -2 \k_k \bounb_{k} & \qquad
\delta\mu(\b_{k}, \b_{l}) = 0
\end{array}
\label{CKusuCobound}
\ee
\be
\delta\mu(\b_k, \e)=0\,.
\label{CKuCobound}
\ee
We shall not need the remaining values of the coboundaries $\delta\mu$ for
${su}_\kbold(N+1)$ or ${u}_\kbold(N+1)$;  each
$\delta\mu$ being a two-cocycle, it must necessarily appear as a
particular case of the most general two-cocycles which are completely
determined by the basic extension coefficients (\ref{suBasicExtCoef}).

The question of whether a general two-cocycle for a CK
algebra in Theorem~\ref{theor3.1} defines a trivial extension
amounts to checking whether it is a coboundary, which will allow to
eliminate the central $\Xi$ term from (\ref{CKusuGenExtCR}). This
may depend on the values of the constants $\k_a$. In fact, the
three types of extensions  behave in three different ways, which
mimics the pattern found in the orthogonal case \cite{Azc.Her.Bue.San:96}:

\begin{itemize}
\item
Type I extensions can be done for all  unitary CK
algebras, since there is not any $\k_a$-dependent
restriction to the basic Type I coefficients
$\cextj_{ab},\ \cextm_{ab}$. However, as seen in
(\ref{CKusuCobound}), these extensions are \emph{always} trivial. A
considerable simplification of all expressions can be gained
if these trivial extensions are simply discarded, as we shall
do from now on. Hence for the \emph{extended}
algebra, the whole block of commutation
relations in (\ref{CKusuCR}) will hold and only those commutators in
(\ref{CKsuCR}) or
(\ref{CKuCR}) may change.

\item
Type II extensions appear also in all  unitary CK
algebras, as there is not any $\k_a$-dependent restriction to the
basic Type II coefficients
$\cexta_{k}$. The triviality of these extensions is
$\k_a$-dependent, and (\ref{CKusuCobound}) shows that the extension
determined by the  coefficient $\cexta_{k}$ is
non-trivial if $\k_{k} = 0$, and trivial otherwise. It is within
this type of extensions that a \emph{pseudoextension} (trivial
extension by a two-coboundary) may become a non-trivial extension
by contraction
\cite{Ald.Azc:85b,Azc.Izq:95}.

\item
Type III extensions behave in a completely different way.
Due to the additional conditions (\ref{ECtaRestrict}) and
(\ref{urestrict}) that Type III extension coefficients must fulfil, some
of them might be necessarily equal to zero. Hence, these extensions do not
exist for all  unitary CK algebras. But those allowed (one
$\cextb_{kl}$ for each pair of vanishing constants $\k_k = \k_l
=0$ and for the (non-special)  unitary case one additional $\cextg_{k}$
for each vanishing constant $\k_k =0$) are
always non-trivial, as the last equation in (\ref{CKusuCobound}) and
(\ref{CKuCobound}) show. Therefore, Type III extensions  do
not appear through the pseudoextension mechanism.
\end{itemize}


\subsection{The second cohomology groups of the unitary CK algebras}
\label{sec4}

If we disregard Type I extensions, which are trivial for all members in
the two CK families of  unitary algebras, the above  results can be
summarised in the following

\begin{theorem}
\label{theor4.1}
The  commutation relations of any central extension
$\overline{su}_\kbold(N+1)$ of the special unitary CK algebra
${su}_\kbold(N+1)$ can be written as the commutation relations in
(\ref{CKusuCR}), together with:
\be
\displaystyle  [\j_{ab},\m_{ab}] = -2\k_{ab}\sum_{s=a+1}^b \b_s +
         \sum_{s=a+1}^b\k_{a\,s-1}\k_{sb}\,\cexta_{s} \Xi  \qquad
[\b_k,\b_l]=\cextb_{kl}\Xi\qquad k<l
\label{CKsuExtCR}
\ee
which will replace those in (\ref{CKsuCR}). The extension is
completely characterised by

\begin{itemize}
\item $N$ Type II coefficients ${\cexta_{k}}$ ($k=1,\dots,N$).
Each of them gives rise to a non-trivial extension if
$\k_k= 0$ and to a trivial one otherwise.

\item $N(N-1)/2$ Type III extension coefficients $\cextb_{kl}$
($k<l$ and
$k,l=1,\dots,N$), satisfying
\be
 \k_k \cextb_{kl}=0\qquad \k_l\cextb_{kl}=0.
\label{eb}
\ee
Thus, $\cextb_{kl}$ must be equal to zero when at least one of the
constants $\k_k,\ \k_l$ is different from zero. When $\cextb_{kl}$ is
non-zero, the extension that it determines is always non-trivial.
\end{itemize}
\end{theorem}

\begin{theorem}
\label{theor4.2}
The  commutation relations of any central extension
$\overline{u}_\kbold(N+1)$ of the unitary CK algebra
${u}_\kbold(N+1)$ can be written as the commutation relations in
the preceding statement, together with
\be
\begin{array}{lll}
[\j_{ab},\e] = 0  & \qquad
[\m_{ab},\e] = 0  & \qquad
[\b_k,\e]=\cextg_{k}\Xi
\label{CKuExtCR}
\end{array}
\ee
which will replace those in (\ref{CKuCR}). In addition to the
extension coefficients ${\cexta_{k}}$ and $\cextb_{kl}$, the
extension is completely characterised by
\begin{itemize}
\item
$N$ Type III coefficients ${\cextg_{k}}$ ($k=1,\dots,N$) satisfying
\be
 \k_k \cextg_{k}=0.
\ee
When $\cextg_k$ is non-zero, the extension that it determines is non-trivial.
\end{itemize}
\end{theorem}

All Type II extensions come from the pseudocohomology mechanism
\cite{Ald.Azc:85b,Azc.Izq:95}. We can write (\ref{CKsuExtCR}) as
\be
[\j_{ab},\m_{ab}]=-2\k_{ab} \sum_{s=a+1}^b \big(  \b_s -
\frac{\cexta_{s}}{2 \k_s} \big) \Xi \quad
\label{TypeIIPseudoExt}
\ee
which is well defined even if any of the $\k_s$
($s=a+1, a+2, \dots, b$) is equal to zero.
This clearly shows that when a given
$\k_s$  is different from zero, the extension
coefficient
$\cexta_{s}$ gives rise to a trivial extension, which can be
removed by the one-cochain
$\mu(B_s)=-\frac{\cexta_s}{2 \k_s}$ (all other coordinates of the
one-cochain being zero). However, when $\k_s$ goes to zero, the
corresponding extension is non-trivial, as the cochain  defined above
diverges, but the term $\k_{ab}/\k_s$
in (\ref{TypeIIPseudoExt}) does not.

In terms of the triangular arrangement for the generators of
${su}_\kbold(N+1)$ (see Fig.~\ref{fig2.1}), it is also  worth remarking
that   Type III extensions only affect the commutators of the Cartan
generators in the outermost `$B$' diagonal, while the Type~II extension
$\cexta_a$  only modifies the commutators of each those pairs
$\{\j_{ij},\m_{ij}\}\equiv\omm_{ij}$ with $i<a \leq j$, {\it i.e.} those pairs
contained inside a rectangle
with left-down corner $\omm_{a-1 a}$.

As a by-product of these results we can give closed expressions for
the dimension of the second cohomology group of any Lie algebra in
the unitary CK families.

\begin{proposition}
\label{prop4.1}
Let ${su}_\kbold(N+1)$ or ${u}_\kbold(N+1)$
be a Lie algebra belonging
to a family of unitary CK algebras, and let $n$ be the
number of coefficients $\k_k$ equal to zero. The dimension of
its second cohomology group is given by
\be
{\mbox{dim}}\,
(H^2({su}_\kbold(N+1),\Re)=n + \frac {n(n-1)}{2} =
  \frac{n(n+1)}2
\label{dimH}
\ee
\be
{\mbox{dim}}\,
(H^2({u}_\kbold(N+1),\Re)=n + \frac {n(n-1)}{2} +n =
  \frac {n(n+3)}2.
\label{uec}
\ee
\end{proposition}

The first term $n$ in the sum of (\ref{dimH}), (\ref{uec}) corresponds to the
central extensions ${\cexta_{k}}$, the second term $\frac
{n(n-1)}{2}$ to the $\cextb_{kl}$ and the third term $n$ in (\ref{uec})
to the central extensions $\cextg_k$. We recall that the analogous
expression for the quasi-orthogonal case is far more complicated,
and depends not only on the number of constants equal to zero, but
also on the detailed arrangement of zeros in the sequence $\k_1,
\dots, \k_N$ \cite{Azc.Her.Bue.San:96}.

As expected for the simple
$su({p,q})$ or the semisimple $u({p,q})$ algebras, which appear within the
two unitary CK families when all
$\k_a\ne 0$, the second cohomology group is trivial. The inhomogeneous
$iu({p,q})$ algebras, appearing in the special unitary family when either
$\k_1=0$ or
$\k_N=0$, with all other constants $\k_a
\neq 0$, have, in any dimension, a single non-trivial extension:
$\cexta_{1}$ when $\k_1=0$ or $\cexta_{N}$ if $\k_N=0$.
The  special unitary flag algebra (when all $\k_a= 0$) has the maximum
number of non-trivial extensions within the special unitary family, that
is, ${N(N+1)}/2$.


\section{Examples}
\label{sec.4}

Let us illustrate the general results of the above section for the
${su}_\kbold(N+1)$ algebras in the three lowest dimensional cases,
$N=1,2,3$. A completely similar discussion can be performed for the
${u}_\kbold(N+1)$ algebras.

\subsection{$\overline{su}_{\k_1}(2)$ }

We simply mention this example for the sake of completeness. The
results for the extensions of ${su}_{\k_1}(2)$ could be also obtained
from those in
\cite{Azc.Her.Bue.San:96}
by using the isomorphism ${su}_{\k_1}(2) \simeq so_{\k_1, +
}(3,
\Re)$ provided by
$\j_{01}/2 \leftrightarrow \Omega_{01}$,
$\m_{01}/2 \leftrightarrow \Omega_{02}$,
$-\b_{1}/2 \leftrightarrow \Omega_{12}$.
The most general extension is defined
by the extension coefficient $\cexta_{1}$
and the non-zero Lie brackets
\be
[\j_{01},\m_{01}] =-2\k_1\b_1+\cexta_{1}\Xi  \qquad
[\j_{01},\b_{1}] =2\m_{01}  \qquad
[\m_{01},\b_{1}] =-2\j_{01}.
\ee
The extension is non-trivial for $\k_1 = 0$ and trivial otherwise, the
triviality being exhibited by the redefinition
\be
\b_1 \to \b_1 - \frac{\cexta_{1}}{2\k_1} \Xi .
\ee

\subsection{$\overline{su}_{\k_1,\k_2}(3)$ }

The most general extended  special unitary CK algebra
$\overline{su}_{\k_1,\k_2}(3)$  has nine generators
$\{\j_{01},\j_{02},\j_{12}, \m_{01},\m_{02},\m_{12},
\b_{1},\b_{2},\Xi\}$, and it is determined by three possible extension
coefficients $\{\cexta_{1}, \cexta_{2}, \cextb_{12}\}$, with
$\k_1\cextb_{12}=\k_2\cextb_{12}=0$.  Their  commutators are:
\be
\begin{array}{lll}
[\j_{01},\j_{02}] =\k_{1}\j_{12} &\qquad
[\j_{01},\j_{12}] =-\j_{02} &\qquad
[\j_{02},\j_{12}] =\k_{2}\j_{01}\cr
[\m_{01},\m_{02}] =\k_{1}\j_{12} &\qquad
[\m_{01},\m_{12}] =\j_{02} &\qquad
[\m_{02},\m_{12}] =\k_{2}\j_{01} \cr
[\j_{01},\m_{02}] =\k_{1}\m_{12} &\qquad
[\j_{01},\m_{12}] =-\m_{02} &\qquad
[\j_{02},\m_{12}] =-\k_{2}\m_{01}\cr
[\m_{01},\j_{02}] =-\k_{1}\m_{12} &\qquad
[\m_{01},\j_{12}] =-\m_{02} &\qquad
[\m_{02},\j_{12}] =\k_{2}\m_{01} \cr
[\j_{01},\b_{1}] =2\m_{01} &\qquad
[\j_{02},\b_{1}] = \m_{02} &\qquad
[\j_{12},\b_{1}] =- \m_{12}\cr
[\j_{01},\b_{2}] =-\m_{01} &\qquad
[\j_{02},\b_{2}] = \m_{02} &\qquad
[\j_{12},\b_{2}] =2\m_{12}\cr
[\m_{01},\b_{1}] =-2\j_{01} &\qquad
[\m_{02},\b_{1}] = -\j_{02} &\qquad
[\m_{12},\b_{1}] =\j_{12}\cr
[\m_{01},\b_{2}] =\j_{01} &\qquad
[\m_{02},\b_{2}] = -\j_{02} &\qquad
[\m_{12},\b_{2}] =-2\j_{12}
\end{array}
\label{qa}
\ee
\be
\begin{array}{l}
[\j_{01},\m_{01}] =-2\k_1\b_1+\cexta_{1}\Xi
\qquad \qquad  [\j_{12},\m_{12}] =-2\k_2\b_2 +   \cexta_{2}\Xi \cr
 [\j_{02},\m_{02}] =\k_2(-2\k_1 \b_1
+\cexta_{1}\Xi)+\k_1(  -2\k_2 \b_2 +   \cexta_{2}\Xi   )\cr
[\b_1,\b_2]=\cextb_{12}\Xi .
\end{array}
\label{CKsuiiiCR}
\ee

The triviality of Type II extensions is governed by the values of the
constants $\k_1,\k_2$. We analyse this problem for each specific CK
algebra within $\overline{su}_{\k_1,\k_2}(3)$. The extension
determined by $\cexta_{1}$ is trivial when $\k_1 \neq 0$, and the
extension determined by $\cexta_{2}$ is trivial when $\k_2 \neq 0$, the
triviality being exhibited by the redefinitions
\be
\b_1 \to \b_1 - \frac{\cexta_{1}}{2\k_1} \Xi  \qquad
\b_2 \to \b_2 - \frac{\cexta_{2}}{2\k_2} \Xi.
\ee
Thus, ${\mbox{dim}}\, (H^2({su}_{\k_1, \k_2}(3),\Re))$  is equal to

\begin{itemize}
\item
0 when both $\k_1,\k_2\ne 0$. Here both
$\cexta_{1}, \cexta_{2}$ produce trivial extensions, and $\cextb_{12}$
must vanish. This case corresponds to the extensions of ${su}(3)$ for
$(\k_1,\k_2)=(+,+)$,  and  ${su}({2,1})$ for
$(\k_1,\k_2)=\{(+,-),(-,+),(-,-)\}$ and the result is in agreement
with Whitehead's lemma, according to which simple algebras have no
non-trivial extensions.

\item
1 for the inhomogeneous unitary algebras ${iu}(2)$ and
${iu}({1,1})$. These algebras appear twice in the CK family, namely
for $\k_1= 0,\ \k_2 \neq 0$ and for $\k_1\neq 0,\ \k_2=0$. In the
first case the only non-trivial extension coefficient is
$\cexta_{1}$ and the extended Lie brackets (\ref{CKsuiiiCR}) reduce to
\be
 [\j_{01},\m_{01}] = \cexta_{1}\Xi  \quad
 [\j_{02},\m_{02}] =\k_2  \cexta_{1}\Xi     \quad
 [\j_{12},\m_{12}] =-2\k_2\b_2  \quad [\b_1,\b_2]=0.
\label{qc}
\ee
The second case  is related to the former one due to the isomorphism
(\ref{PolarityIsom}). Here there is a single non-trivial extension
coefficient $\cexta_{2}$ and the extended Lie brackets are
\be
[\j_{01},\m_{01}] =-2\k_1\b_1  \quad
[\j_{02},\m_{02}] =\k_1  \cexta_{2}\Xi   \quad
[\j_{12},\m_{12}] = \cexta_{2}\Xi  \quad
[\b_1,\b_2]=0.
\label{qd}
\ee

\item
3 for the special unitary flag algebra ${su}_{0,0}(3)$
when $\k_1=\k_2= 0$. The three extensions are  non-trivial
\be
[\j_{01},\m_{01}] =\cexta_{1}\Xi  \quad
[\j_{02},\m_{02}] =0  \quad
[\j_{12},\m_{12}] = \cexta_{2}\Xi  \quad
[\b_1,\b_2]=\cextb_{12}\Xi.
\label{qe}
\ee
\end{itemize}

\subsection{$\overline{su}_{\k_1,\k_2, \k_3}(4)$ }

We consider now the extensions $\overline{su}_{\k_1,\k_2,\k_3}(4)$ of
the CK algebra ${su}_{\k_1,\k_2,\k_3}(4)$. There are six possible
basic extension coefficients, $\{\cexta_{1},\cexta_{2},\cexta_{3},
\cextb_{12},\cextb_{13},\cextb_{23}\}$, which must satisfy the
conditions
\be
\k_1 \cextb_{12}=\k_2 \cextb_{12}=0   \qquad
\k_1 \cextb_{13}=\k_3 \cextb_{13}=0   \qquad
\k_2 \cextb_{23}=\k_3 \cextb_{23}=0,
\ee
and the Lie brackets of the extension are given by the non-extended
ones in (\ref{CKusuCR}) and by the extended ones
\be
\begin{array}{l}
[\j_{01},\m_{01}] =-2\k_1\b_1+\cexta_{1}\Xi\cr
[\j_{02},\m_{02}] =\k_2(-2\k_1 \b_1 + \cexta_{1}\Xi)+\k_1(
-2\k_2 \b_2 +   \cexta_{2}\Xi   )\cr
[\j_{03},\m_{03}] =\k_2\k_3(-2\k_1 \b_1 + \cexta_{1}\Xi)
 +\k_1\k_3( -2\k_2 \b_2 +   \cexta_{2}\Xi   ) \cr
\qquad\qquad\qquad
 +\k_1\k_2( -2\k_3 \b_3 +   \cexta_{3}\Xi   )\cr
[\j_{12},\m_{12}] =-2\k_2\b_2 +   \cexta_{2}\Xi\cr
[\j_{13},\m_{13}] =\k_3(-2\k_2 \b_2 + \cexta_{2}\Xi)+\k_2(
-2\k_3 \b_3 +   \cexta_{3}\Xi   )\cr
[\j_{23},\m_{23}] =-2\k_3\b_3 +   \cexta_{3}\Xi\cr
[\b_1,\b_2]=\cextb_{12}\Xi \qquad [\b_1,\b_3]=\cextb_{13}\Xi \qquad
[\b_2,\b_3]=\cextb_{23}\Xi .
\end{array}
\label{qf}
\ee

The results for each one of the 27 CK algebras
$\overline{su}_{\k_1,\k_2,\k_3}(4)$ are displayed in Table~\ref{table4.1}.
The columns in this Table show, in this order, the number of coefficients
$\k_a$ set equal to zero (number of contractions), the centrally
extended Lie algebras, the signs $+, -, 0$ of each
coefficient $(\k_1,\k_2,\k_3)$  together with the  non-trivial central
extensions  allowed for the algebra with these signs for the coefficients,
and finally, the dimension of the second cohomology group as a sum of
the number of non-trivial extensions of Types II and III, coming
respectively from the coefficients $\cexta_{k}$ and $\cextb_{kl}$. In the
table $+$ ($-$) denotes a positive (negative) $\k_a$ coefficient which could be
rescaled to $1$ ($-1$).

\begin{table}
\begin{center}
\begin{tabular}{|l|l|l|l|}
\hline
\multicolumn{1}{|c|}{$\!$\#$\!$} &\multicolumn{1}{c|}{Extended
algebra}& \multicolumn{1}{c|}{(CK constants) [Non-trivial
extensions]$\!$}& \multicolumn{1}{c|}{$\!$dim$H^2\!\!$}\\[0.1cm]
\hline
0&$\overline{su}({4})$&$(+,+,+)$&  0\\[0.1cm]
 &$\overline{su}({3,1})$&$(-,+,+),(-,-,+),(+,+,-),(+,-,-)$   & \cr
 &$\overline{su}({2,2})$ &$(+,-,+),(-,+,-),(-,-,-)$& \\[0.1cm]
\hline
1 &$\overline{iu}({3})$&($0,+,+$) $[\cexta_{1}]$
or ($+,+,0$) $[\cexta_{3}]$  & 1+0\\[0.1cm]
  &$\overline{iu}({2,1})$&$(0,-,+),(0,+,-),(0,-,-)$  $[\cexta_{1}]$
or&  \\[0.1cm]
 & &$(+,-,0),(-,+,0),(-,-,0)$  $[\cexta_{3}]$&  \\[0.1cm]
  &$\overline{t}_8(u({2}) \oplus u(1) \oplus u(2))$&$(+,0,+)$
$[\cexta_{2}]$ &  \\[0.1cm]
  &$\overline{t}_8(u({2}) \oplus u(1) \oplus  u({1,1}))$&$(+,0,-),(-,0,+)$
$[\cexta_{2}]$  &  \\[0.1cm]
  &$\overline{t}_8(u({1,1}) \oplus u(1) \oplus  u({1,1}))$&$(-,0,-)$
$[\cexta_{2}]$ &  \\[0.1cm]
\hline
2 & &($0,0,+$) $[\cexta_{1},\cexta_{2},\cextb_{12}]$
or ($+,0,0$) $[\cexta_{2},\cexta_{3},\cextb_{23}]$   & 2+1\\[0.1cm]
  & &($0,0,-$) $[\cexta_{1},\cexta_{2},\cextb_{12}]$
or ($-,0,0$) $[\cexta_{2},\cexta_{3},\cextb_{23}]$   &  \\[0.1cm]
 & &($0,+,0$) $[\cexta_{1},\cexta_{3},\cextb_{13}]$ &  \\[0.1cm]
 & &($0,-,0$) $[\cexta_{1},\cexta_{3},\cextb_{13}]$ &  \\[0.1cm]
\hline
3 &Flag algebra&($0,0,0$)   $[\cexta_{1},\cexta_{2},\cexta_{3},
\cextb_{12},\cextb_{13},\cextb_{23}]$  & 3+3\\[0.1cm]
\hline
\end{tabular}
\end{center}
\caption{Non-trivial central extensions
$\overline{su}_{\k_1,\k_2,\k_3}(4)$.}
\label{table4.1}
\end{table}

\section{Conclusions and outlook}

We restrict here to a couple of remarks. First, the pattern of three
types of extensions behaving under contractions in three different ways,
first found for the quasi-orthogonal family \cite{Azc.Her.Bue.San:96}, appears
also in the quasi-unitary
case. This seems likely to be a general phenomenon, not restricted to a single
family of contractions of some Lie algebras. The analysis of the extensions for
the
third CK main series of algebras, which embraces the symplectic
$sp(p,q)$ in
the $C_l$ series and their contractions, would be required to complete the
study of the relationships between cohomology and contractions undertaken
in \cite{Azc.Her.Bue.San:96} and continued in this paper.
These algebras can be adequately realised by quaternionic antihermitian
matrices, or, alternatively, by quaternionic antihermitian traceless matrices
plus
the Lie algebra of derivations of the quaternion
division algebra. Work in this area is in progress. Second, as compared to the
quasi-orthogonal case, the
quasi-unitary algebras have a comparatively smaller set of
extensions, whose description in terms of the values taken by the CK constants
$\k_a$ is straightforward. The suitability of a CK approach to the study of the
central extensions of a complete family is therefore put forward more clearly
than in the orthogonal case.
While the ordinary inhomogeneous orthogonal algebras
$iso(p,q)$ associated to the real orthogonal $N=p+q$
dimensional flat spaces have
non-trivial extensions only in the case $N=2$, the algebras $iu(p,q)$
associated to
the complex pseudo-Euclidean hermitian flat spaces have a
single non-trivial extension, in any dimension. The
relevance of this fact in relation with the classical limit of quantum
mechanics
will be discussed elsewhere.


\section*{Acknowledgements}
The authors wish to acknowledge J. A. de Azc\'arraga for his comments on the
manuscript.
This research has been partially supported by the Spanish DGES
projects PB96--0756, PB94--1115
from the Ministerio de Educaci\'on y Cultura and by Junta de Castilla y
Le\'on (Projects CO1/396 and CO2/297). J.C.P.B.
wishes to thank an FPI grant from the Ministerio de Educaci\'on y
Cultura and the CSIC.


\section*{Appendix: The general solution to the Jacobi identities}
\setcounter{equation}{0}
\renewcommand{\theequation}{A.\arabic{equation}}

In order to get the general solution of the set of linear equations
determining the possible extensions of the  unitary CK
algebras, we first introduce a suitable notation for the central
extension coefficients, which is `adapted' to the structure of the
algebras $su_{\kbold}(N+1)$  (\ref{CKusuCR})--(\ref{CKsuCR})  and
$u_{\kbold}(N+1)$ (\ref{CKusuCR})--(\ref{CKuCR})
whose basic generators come  naturally divided in either three
or four `kinds' $\j_{ab},\ \m_{ab},\ \b_{k},\ \e$. The symbol
corresponding to $\cext(X, Y)$ will have one or two letters taken
from $j, m, b, i$, determined by the kind of the basis generators $X,\
Y$. To this symbol we append two groups of indices, each coming from
those of the corresponding generators.  The complete list of all
extension coefficients as written in this notation is
\be
\begin{array}{llll}
 \jj_{ab,de}&\qquad \mm_{ab,de}&\qquad \jm_{ab,de} &\qquad \mj_{ab,de}\cr
 \jb_{ab,k} &\qquad\mb_{ab,k} &\qquad\bb_{k,l} &\qquad\jjmm_{ab} \cr
 \je_{ab} &\qquad \me_{ab}&\qquad \bee_l &\qquad
\end{array}
\label{EC}
\ee
where we implicitly assume $a<b,\ d<e,\ a,b,d,e=0, \dots N, \ k<l,
\ k, l=1, \dots N$. We remark that $\jm,\ \mj,\ \jb,\ \mb,\ \je,\ \me,\
\bee$ are single, unbreakable symbols, and are not products. In the
course of the derivation we will find useful to sort these
coefficients into several subsets, as follows

\begin{itemize}
\item Coefficients
$\jj_{ab,de},\ \mm_{ab,de},\ \jm_{ab,de},\ \mj_{ab,de}$
involving \emph{four} different indices. If we write these four indices
as $a<b<c<d$ the coefficients are
\be
\begin{array}{llll}
 \jj_{ab,cd}&\qquad \mm_{ab,cd}&\qquad \jm_{ab,cd} &\qquad \mj_{ab,cd}\cr
 \jj_{ac,bd}&\qquad \mm_{ac,bd}&\qquad \jm_{ac,bd} &\qquad \mj_{ac,bd}\cr
 \jj_{ad,bc}&\qquad \mm_{ad,bc}&\qquad \jm_{ad,bc} &\qquad \mj_{ad,bc}\cr
\end{array}
\label{ECa}
\ee
\item Coefficients
$\jj_{ab,de}, \mm_{ab,de}, \jm_{ab,de}, \mj_{ab,de}$ involving
\emph{three} different indices. If we write the three indices as
$a<b<c$ these coefficients are
\be
\begin{array}{llll}
 \jj_{ab,ac}&\qquad \mm_{ab,ac}&\qquad \jm_{ab,ac} &\qquad \mj_{ab,ac}\cr
 \jj_{ab,bc}&\qquad \mm_{ab,bc}&\qquad \jm_{ab,bc} &\qquad \mj_{ab,bc}\cr
 \jj_{ac,bc}&\qquad \mm_{ac,bc}&\qquad \jm_{ac,bc} &\qquad \mj_{ac,bc}\cr
\end{array}
\label{ECb}
\ee
\item Coefficients $\jjmm_{ab}$
involving \emph{two} different indices
\be
\begin{array}{l}
 \jjmm_{ab} \cr
\end{array}
\label{ECc}
\ee
\item Coefficients $\jb_{ab, i}, \mb_{ab, i}$ with \emph{two}
different indices $a<b$ and a third index $i \in \{a,a+1,b,b+1\}$
\be
\begin{array}{ll}
 \jb_{ab,i} &\qquad \mb_{ab,i}
\end{array}
\label{ECd}
\ee
\item Coefficients $\jb_{ab, j}, \mb_{ab, j}$ with \emph{two}
different indices $a<b$ and a third index $j \notin \{a,a+1,b,b+1\}$
\be
\begin{array}{ll}
 \jb_{ab,j} &\qquad \mb_{ab,j}
\end{array}
\label{ECe}
\ee
\item Coefficients $\bb_{k,l}$ with two different indices $k<l$
\be
\begin{array}{l}
 \bb_{k,l} \cr
\end{array}
\label{ECf}
\ee
\item Coefficients $\je_{ab}$ and $\me_{ab}$ with \emph{two}
different indices $a<b$
\be
\begin{array}{ll}
 \je_{ab}  &\qquad \me_{ab}
\end{array}
\label{ECg}
\ee
\item Coefficients $\bee_{l}$ with a \emph{single} index
\be
\begin{array}{ll}
 \bee_{l}\ .
\end{array}
\label{ECh}
\ee
\end{itemize}
The Lie brackets of the extended CK algebra
$\overline{su}_\kbold(N+1)$  and
$\overline{u}_\kbold(N+1)$   read
\be
\begin{array}{ll}
[\j_{ab},\j_{ac}] =\k_{ab}\j_{bc}+\jj_{ab,ac}\Xi &\quad
[\m_{ab},\m_{ac}] =\k_{ab}\j_{bc}+\mm_{ab,ac}\Xi
\\{}
[\j_{ab},\j_{bc}] =-\j_{ac} -\jj_{ab,bc}\Xi &\quad
[\m_{ab},\m_{bc}] =\j_{ac}+\mm_{ab,bc}\Xi
\\{}
[\j_{ac},\j_{bc}] =\k_{bc}\j_{ab} + \jj_{ac,bc}\Xi  &\quad
[\m_{ac},\m_{bc}] =\k_{bc}\j_{ab} + \mm_{ac,bc}\Xi
\\{}
[\j_{ab},\j_{de}]= \jj_{ab,de}\Xi &\quad
[\m_{ab},\m_{de}] =\mm_{ab,de}\Xi
\\[0.3cm]
[\j_{ab},\m_{ac}] =\k_{ab}\m_{bc}+\jm_{ab,ac} \Xi &\quad
[\m_{ab},\j_{ac}] =-\k_{ab}\m_{bc} - \mj_{ab,ac}\Xi
\\{}
[\j_{ab},\m_{bc}] =-\m_{ac}- \jm_{ab,bc} \Xi &\quad
[\m_{ab},\j_{bc}] =-\m_{ac} - \mj_{ab,bc} \Xi
\\{}
[\j_{ac},\m_{bc}] =-\k_{bc}\m_{ab} - \jm_{ac,bc}\Xi &\quad
[\m_{ac},\j_{bc}] =\k_{bc}\m_{ab} + \mj_{ac,bc} \Xi
\\{}
[\j_{ab},\m_{de}] =\jm_{ab,de}\Xi &\quad
[\m_{ab},\j_{de}] =\mj_{ab,de}\Xi
\end{array}
\label{CKsuExtA}
\ee
\be
\begin{array}{ll}
[\j_{ab},\b_a]=-\m_{ab}-\jb_{ab,a}\Xi &
[\m_{ab},\b_a]=\j_{ab}+\mb_{ab,a} \cr
[\j_{ab},\b_{a+1}]=\m_{ab}+\jb_{ab,a+1}\Xi &
[\m_{ab},\b_{a+1}]=-\j_{ab}-\mb_{ab,a+1}\Xi \quad  b\ge a+2\cr
[\j_{a\,a+1},\b_{a+1}]=2\m_{a\,a+1}+2\jb_{a\,a+1,a+1}\Xi
& [\m_{a\,a+1},\b_{a+1}]=-2\j_{a\,a+1}-2\mb_{a\,a+1,a+1} \Xi\cr
[\j_{ab},\b_{b}]=\m_{ab}+\jb_{ab,b}\Xi&
[\m_{ab},\b_{b}]=-\j_{ab}-\mb_{ab,b}\Xi \quad b\ge a+2\cr
[\j_{ab},\b_{b+1}]=-\m_{ab}-\jb_{ab,b+1}\Xi&
[\m_{ab},\b_{b+1}]=\j_{ab}+\mb_{ab,b+1}\Xi \cr
[\j_{ab},\b_{j}]=\jb_{ab,j}\Xi &
[\m_{ab},\b_{j}]=\mb_{ab,j}\Xi
\end{array}
\label{CKsuExtB}
\ee
\be
[\j_{ab},\m_{ab}]=-2\k_{ab}\sum_{s=a+1}^b \b_s + \jjmm_{ab} \Xi
\qquad\qquad
[\b_k,\b_l]=\bb_{k,l} \Xi
\label{bf}
\ee
\be
\begin{array}{lll}
[\j_{ab},\e] = \je_{ab} \Xi & \qquad
[\m_{ab},\e] = \me_{ab} \Xi & \qquad
[\b_l,\e] = \bee_l \Xi
\label{CKuExtC}
\end{array}
\ee
where as indicated before, the relations $a<b<c$, $a<d$, $d<e$,   $j
\notin\{a,a+1,b,b+1\}$,
$k<l$ for the indices $a,b,c,d,e=0, \dots N, \ j,k,l=1,\dots, N$
and $a, b, d, e$ are all different, will be assumed without saying.

Our strategy here will be to enforce the complete set of Jacobi
identities  first for $su_{\kbold}(N+1)$ and then for $u_{\kbold}(N+1)$, in
a carefully selected order which actually allows to explicitly
solve the rather large set of linear equations. The first stage
will be to identify many extension coefficients which are forced to
vanish; the remaining Jacobi equations will drastically simplify
and will either produce relations allowing to express certain
\emph{derived} extension coefficients in terms of the so-called
\emph{basic} ones, or further relations to be satisfied by the basic
extension coefficients.

To begin with, we show that \emph{all} coefficients in
(\ref{ECa}) vanish.  Denoting by $\{X,Y,Z\}$ the Jacobi
identity for the generators $X$,
$Y$ and $Z$, we display several choices for them
and the equations ensuing from these choices:
\be
\begin{array}{ll}
\{\j_{ab},\m_{cd},\b_d\}:&\ \jj_{ab,cd}=0\cr
\{\j_{ab},\m_{cd},\b_b\}:&\ \mm_{ab,cd}=0\cr
\{\j_{ab},\j_{cd},\b_d\}:&\ \jm_{ab,cd}=0\cr
\{\j_{ab},\j_{cd},\b_b\}:&\ \mj_{ab,cd}=0
\end{array}
\label{JEa}
\ee
\be
\begin{array}{ll}
\{\j_{ad},\m_{bc},\b_c\}:&\ \jj_{ad,bc}=0\cr
\{\m_{ad},\j_{bc},\b_c\}:&\ \mm_{ad,bc}=0\cr
\{\j_{ad},\j_{bc},\b_c\}:&\ \jm_{ad,bc}=0\cr
\{\m_{ad},\m_{bc},\b_c\}:&\ \mj_{ad,bc}=0
\end{array}
\label{JEb}
\ee
\be
\begin{array}{lll}
\{\j_{ab},\j_{bc},\j_{bd}\}:&\quad
\k_{bc}\jj_{ab,cd}+\jj_{ac,bd}-\jj_{ad,bc}=0\cr
\{\j_{ab},\m_{bc},\m_{bd}\}:&\quad
\k_{bc}\jj_{ab,cd}+\mm_{ac,bd}-\mm_{ad,bc}=0\cr
\{\j_{ab},\j_{bc},\m_{bd}\}:&\quad
\k_{bc}\jm_{ab,cd}+\jm_{ac,bd}-\mj_{ad,bc}=0\cr
\{\j_{ab},\m_{bc},\j_{bd}\}:&\quad
\k_{bc}\jm_{ab,cd}-\mj_{ac,bd}+\jm_{ad,bc}=0 .\cr
\end{array}
\label{JEc}
\ee
By substituting (\ref{JEa}) and (\ref{JEb}) in (\ref{JEc}), we find
that \emph{all} coefficients in (\ref{ECa})  are necessarily equal to
zero. From now on, substitution of the already known information in
further equations will be automatically assumed.

The coefficients  in (\ref{ECe}) turn out to be also equal to zero:
\be
\{\m_{ab},\b_{b},\b_j\}: \ \jb_{ab,j}=0 \quad
\{\j_{ab},\b_{b},\b_j\}: \ \mb_{ab,j}=0 \quad
j\notin \{a,a+1,b,b+1\} .
\label{cd}
\ee

Now we look for equations involving the  coefficients $\bb_{k,l}$
in (\ref{ECf}). We find:
\be
\begin{array}{llll}
\{\j_{a\, a+1},\m_{a\, a+1},\b_{k}\}:&\
\k_{a\,a+1}\bb_{k,a+1}=0&\quad 1\le k\le a&\quad a=1,\dots,N-1\cr
\{\j_{b-1\, b},\m_{b-1\, b},\b_{l}\}:&\
\k_{b-1\,b}\bb_{b,l}=0&\quad b+1\le l\le N&\quad b=1,\dots,N-1
\end{array}
\label{JEf}
\ee
so the $N(N-1)/2$ coefficients of the type $\bb_{k,l}$ might be
different from zero. We denote them as
\be
\cextb_{kl}:=\bb_{k,l}
\label{cee}
\ee
and from  (\ref{JEf}) they must fulfil two additional conditions
\be
\k_k \cextb_{kl}=0\qquad \k_l \cextb_{kl}=0.
\label{cff}
\ee

We now look for Jacobi identities involving the  extension
coefficients in (\ref{ECd}):
\be
\begin{array}{llll}
\{\m_{ab},\b_{a},\b_{a+1}\}:&\jb_{ab,a+1}=\jb_{ab,a}&\
\{\j_{ab},\b_{a},\b_{a+1}\}:&\mb_{ab,a+1}=\mb_{ab,a}\cr
\{\m_{ab},\b_{a},\b_{b}\}:&\jb_{ab,b}=\jb_{ab,a}&\
\{\j_{ab},\b_{a},\b_{b}\}:&\mb_{ab,b}=\mb_{ab,a}\cr
\{\m_{ab},\b_{a},\b_{b+1}\}:&\jb_{ab,b+1}=\jb_{ab,a}&\
\{\j_{ab},\b_{a},\b_{b+1}\}:&\mb_{ab,b+1}=\mb_{ab,a}\cr
\{\m_{ab},\b_{a+1},\b_{b}\}:&\jb_{ab,a+1}=\jb_{ab,b}&\
\{\j_{ab},\b_{a+1},\b_{b}\}:&\mb_{ab,a+1}=\mb_{ab,b}\cr
\{\m_{ab},\b_{a+1},\b_{b+1}\}:&\jb_{ab,a+1}=\jb_{ab,b+1}&\
\{\j_{ab},\b_{a+1},\b_{b+1}\}:&\mb_{ab,a+1}=\mb_{ab,b+1}\cr
\{\m_{ab},\b_{b},\b_{b+1}\}:&\jb_{ab,b}=\jb_{ab,b+1}&\
\{\j_{ab},\b_{b},\b_{b+1}\}:&\mb_{ab,b}=\mb_{ab,b+1}
\end{array}
\label{cf}
\ee
which hold no matter of either $b=a+1$ or $b\ne a+1$. These
equations show that
\be
\begin{array}{l}
\jb_{ab,a}=\jb_{ab,a+1}=\jb_{ab,b}=\jb_{ab,b+1}\cr
\mb_{ab,a}=\mb_{ab,a+1}=\mb_{ab,b}=\mb_{ab,b+1}
\end{array}
\label{cg}
\ee
and therefore these coefficients  only depend on the first pair of
indices. These common values must be considered as another set of
\emph{basic} coefficients
\be
\cextj_{ab}:=\jb_{ab,i}  \qquad \cextm_{ab}:=\mb_{ab,i} \qquad
i\in\{a,a+1,b,b+1\}.
\ee

Now we consider Jacobi identities leading to equations which involve
the coefficients in  (\ref{ECb}), those $\jj_{ab,de},\ \mm_{ab,de},\
\jm_{ab,de},\ \mj_{ab,de}$ with \emph{three} different indices. This
is the most tedious part of the process, due to the need of paying
minute attention to the index ranges.  Let us first look for
equations involving the coefficients with indices $\{ab, bc\}$, which
appear in the middle line of (\ref{ECb})
\be
\begin{array}{lll}
\{\j_{ab},\m_{bc},\b_{c+1}\}:&\quad \jj_{ab,bc}=\cextm_{ac}&\quad c<N
\cr
\{\j_{ab},\m_{bN},\b_{N}\}:&\quad\jj_{ab,bN}=\cextm_{aN}&\quad b<N-1\cr
\{\j_{a\,N-1},\m_{N-1\,N},\b_{a}\}:
&\quad\mm_{a\,N-1,N-1\,N}=\cextm_{aN}&\quad
a>0\cr
\{\j_{0\,N-1},\m_{N-1\,N},\b_{1}\}:
&\quad\mm_{0\,N-1,N-1\,N}=\cextm_{0N}&\cr
 \{\m_{ab},\j_{bc},\b_{c+1}\}:&\quad\mm_{ab,bc}=\cextm_{ac}&\quad c<N
\cr
\{\m_{ab},\j_{bN},\b_{N}\}:&\quad\mm_{ab,bN}=\cextm_{aN}&\quad b<N-1\cr
\{\m_{aN-1},\j_{N-1N},\b_{a}\}:&\quad\jj_{aN-1,N-1N}=\cextm_{aN}
&\quad a>0\cr
\{\m_{0N-1},\j_{N-1N},\b_{1}\}:&\quad \jj_{0N-1,N-1N}=\cextm_{0N}
\end{array}
\label{ch}
\ee
\be
\begin{array}{lll}
\{\j_{ab},\j_{bc},\b_{c+1}\}:&\quad \jm_{ab,bc}=\cextj_{ac}&\quad c<N
\cr
\{\j_{ab},\j_{bN},\b_{N}\}:&\quad\jm_{ab,bN}=\cextj_{aN}&\quad b<N-1\cr
\{\j_{a\,N-1},\j_{N-1\,N},\b_{a}\}:
&\quad\mj_{a\,N-1,N-1\,N}=\cextj_{aN}&\quad
a>0\cr
\{\j_{0\,N-1},\j_{N-1\,N},\b_{1}\}:
&\quad\mj_{0\,N-1,N-1\,N}=\cextj_{0N}&\cr
 \{\m_{ab},\m_{bc},\b_{c+1}\}:&\quad\mj_{ab,bc}=\cextj_{ac}&\quad c<N
\cr
\{\m_{ab},\m_{bN},\b_{N}\}:&\quad\mj_{ab,bN}=\cextj_{aN}&\quad b<N-1\cr
\{\m_{aN-1},\m_{N-1N},\b_{a}\}:&\quad\jm_{aN-1,N-1N}=\cextj_{aN}
&\quad a>0\cr
\{\m_{0N-1},\m_{N-1N},\b_{1}\}:&\quad \jm_{0N-1,N-1N}=\cextj_{0N},
\end{array}
\label{ci}
\ee
so in all cases, and no matter on the value of the middle index $b$,
we have
\be
\jj_{ab,bc}=\mm_{ab,bc}=\cextm_{ac}  \qquad
\jm_{ab,bc}=\mj_{ab,bc}=\cextj_{ac}.
\label{cii}
\ee

For the coefficients in the first line of (\ref{ECb}) we obtain that
\be
\begin{array}{lll}
\{\j_{ab},\m_{ac},\b_{c+1}\}:
&\quad \jj_{ab,ac}=\k_{ab}\cextm_{bc}&\quad c<N
\cr
\{\j_{ab},\m_{aN},\b_{N}\}:&\quad\jj_{ab,aN}=\k_{ab}\cextm_{bN}&\quad
b<N-1\cr
\{\j_{a\,N-1},\m_{aN},\b_{N-1}\}:
&\quad\mm_{a\,N-1,aN}=\k_{aN-1}\cextm_{N-1\,N}&\quad
a<N-2\cr
 \{\m_{ab},\j_{ac},\b_{c+1}\}:
&\quad\mm_{ab,ac}=\k_{ab}\cextm_{bc}&\quad c<N
\cr
\{\m_{ab},\j_{aN},\b_{N}\}:&\quad\mm_{ab,aN}=\k_{ab}\cextm_{bN}&\quad
b<N-1\cr
\{\m_{aN-1},\j_{aN},\b_{N-1}\}:
&\quad\jj_{aN-1,aN}=\k_{aN-1}\cextm_{N-1\,N}
&\quad a<N-2\cr
\multicolumn{3}{l}
{\{\j_{N-2\,N-1},\m_{N-2\,N},\b_{N-1}\}:}
\cr
\multicolumn{3}{l}
{\qquad \jj_{N-2\,N-1,N-2\,N}+\k_{N-2\,N-1}\cextm_{N-1\,N}
-2\mm_{N-2\,N-1,N-2\,N}=0}
\cr
\multicolumn{3}{l}{
\{\m_{N-2\,N-1},\j_{N-2\,N},\b_{N-1}\}:}
\cr
\multicolumn{3}{l}{
\qquad -2\jj_{N-2\,N-1,N-2\,N}+\k_{N-2\,N-1}\cextm_{N-1\,N}
+\mm_{N-2\,N-1,N-2\,N}=0 }
\end{array}
\label{cj}
\ee
\be
\begin{array}{lll}
\{\j_{ab},\j_{ac},\b_{c+1}\}:
&\quad \jm_{ab,ac}=\k_{ab}\cextj_{bc}&\quad c<N
\cr
\{\j_{ab},\j_{aN},\b_{N}\}:&\quad\jm_{ab,aN}=\k_{ab}\cextj_{bN}&\quad
b<N-1\cr
\{\j_{a\,N-1},\j_{aN},\b_{N-1}\}:
&\quad\mj_{a\,N-1,aN}=\k_{aN-1}\cextj_{N-1\,N}&\quad
a<N-2\cr
 \{\m_{ab},\m_{ac},\b_{c+1}\}:
&\quad\mj_{ab,ac}=\k_{ab}\cextj_{bc}&\quad c<N
\cr
\{\m_{ab},\m_{aN},\b_{N}\}:&\quad\mj_{ab,aN}=\k_{ab}\cextj_{bN}&\quad
b<N-1\cr
\{\m_{aN-1},\m_{aN},\b_{N-1}\}:
&\quad\jm_{aN-1,aN}=\k_{aN-1}\cextj_{N-1\,N}
&\quad a<N-2\cr
\multicolumn{3}{l}{
\{\j_{N-2\,N-1},\j_{N-2\,N},\b_{N-1}\}:
}
\cr
\multicolumn{3}{l}{
\qquad \jm_{N-2\,N-1,N-2\,N}+\k_{N-2\,N-1}\cextj_{N-1\,N}
+2\mj_{N-2\,N-1,N-2\,N}=0
}
\cr
\multicolumn{3}{l}{
\{\m_{N-2\,N-1},\m_{N-2\,N},\b_{N-1}\}:
}
\cr
\multicolumn{3}{l}{
\qquad -2\jm_{N-2\,N-1,N-2\,N}+\k_{N-2\,N-1}\cextj_{N-1\,N}
-\mj_{N-2\,N-1,N-2\,N}=0}.
\end{array}
\label{ck}
\ee
These equations are summarised in
\be
\jj_{ab,ac}=\mm_{ab,ac}=\k_{ab}\cextm_{bc}  \qquad
\jm_{ab,ac}=\mj_{ab,ac}=\k_{ab}\cextj_{bc}
\label{cl}
\ee
so again these are derived extension coefficients, expressible in
terms of $\cextm_{bc}$ and $\cextj_{bc}$.

For the coefficients in the third line of (\ref{ECb}) with indices
$\{ac,bc\}$ we get
\be
\begin{array}{lll}
\{\j_{ac},\m_{bc},\b_{a}\}:&
\quad \mm_{ac,bc}=\k_{bc}\cextm_{ab}&\quad a>0
\cr
\{\j_{0c},\m_{bc},\b_{1}\}:&\quad\mm_{0c,bc}=\k_{bc}\cextm_{0b}&\quad
b>1\cr
\{\j_{0c},\m_{1c},\b_{2}\}:
&\quad\jj_{0c,1c}=\k_{1c}\cextm_{01}&\quad
c>2\cr
 \{\m_{ac},\j_{bc},\b_{a}\}:&
\quad\jj_{ac,bc}=\k_{bc}\cextm_{ab}&\quad a>0
\cr
\{\m_{0c},\j_{bc},\b_{1}\}:&\quad\jj_{0c,bc}=\k_{bc}\cextm_{0b}&\quad
b>1\cr
\{\m_{0c},\j_{1c},\b_{2}\}:&\quad\mm_{0c,1c}=\k_{1c}\cextm_{01}
&\quad c>2\cr
\{\j_{02},\m_{12},\b_{2}\}:&\quad
-2 \jj_{02,12}+\k_{12}\cextm_{01}
+\mm_{02,12}=0\cr
\{\m_{02},\j_{12},\b_{2}\}: &\quad
  2\mm_{02,12}-\k_{12}\cextm_{01}
-\jj_{02,12}=0
\end{array}
\label{cm}
\ee
\be
\begin{array}{lll}
\{\j_{ac},\j_{bc},\b_{a}\}:&\quad \mj_{ac,bc}=
\k_{bc}\cextj_{ab}&\quad a>0
\cr
\{\j_{0c},\j_{bc},\b_{1}\}:&\quad\mj_{0c,bc}=\k_{bc}\cextj_{0b}&\quad
b>1\cr
\{\j_{0c},\j_{1c},\b_{2}\}:
&\quad\jm_{0c,1c}=\k_{1c}\cextj_{01}&\quad
c>2\cr
 \{\m_{ac},\m_{bc},\b_{a}\}:&\quad\jm_{ac,bc}=
\k_{bc}\cextj_{ab}&\quad a>0
\cr
\{\m_{0c},\m_{bc},\b_{1}\}:&\quad\jm_{0c,bc}=\k_{bc}\cextj_{0b}&\quad
b>1\cr
\{\m_{0c},\m_{1c},\b_{2}\}:&\quad\mj_{0c,1c}=\k_{1c}\cextj_{01}
&\quad c>2\cr
\{\j_{02},\j_{12},\b_{2}\}:&\quad
2 \jm_{02,12} - \k_{12}\cextj_{01}
- \mj_{02,12}=0\cr
\{\m_{02},\m_{12},\b_{2}\}: &\quad
  - \jm_{02,12}-\k_{12}\cextj_{01}
+2\mj_{02,12}=0 .
\end{array}
\label{cn}
\ee
These equations lead to
\be
\jj_{ac,bc}=\mm_{ac,bc}=\k_{bc}\cextm_{ab} \qquad
\jm_{ac,bc}=\mj_{ac,bc}=\k_{bc}\cextj_{ab}\ ,
\label{co}
\ee
so these coefficients are also derived.

Finally we look for equations involving the coefficients in
(\ref{ECc}), this is $\jjmm_{ac}$. Whenever there exists an index
$b$ between $a$ and $c$, the choice
\be
\{\j_{ab},\j_{ac},\m_{bc}\}:\quad
\jjmm_{ac}=\k_{bc}\jjmm_{ab}+\k_{ab}\jjmm_{bc}
\label{cp}
\ee
leads to an expression for $\jjmm_{ac}$ in terms of $\jjmm_{ab}$ and
$\jjmm_{bc}$. By iterating while possible, we find that the
coefficients   $\jjmm_{ac}$ with $a$ and $c$ not contiguous can be
written in terms of $\jjmm_{ab}$ with $a$ and $b$ contiguous. These
must be considered as basic ones
\be
\cexta_{k}:= \jjmm_{k-1\,k}\qquad k=1,\dots,N \ ,
\label{cq}
\ee
and the remaining coefficients in (\ref{ECc}) are given, recalling that
$\k_{ii}\equiv 1$, by
\be
\jjmm_{ab}=\sum_{s=a+1}^b\k_{a\,s-1}\k_{sb}\,\cexta_{s}
\qquad b\ge a+2.
\ee

As far as $su_{\kbold}(N+1)$ is concerned, the final step in this
process is to ascertain that there is no  any relation for the
extensions coefficients further to the ones yet considered. It can
be checked that \emph{all} remaining Jacobi equations
involving the generators $\j_{ab}, \m_{ab}, \b_l$ are
identically satisfied, so the process has indeed terminated.

Now we deal with the $u_{\kbold}(N+1)$ case; as Jacobi equations
involving $\j_{ab}, \m_{ab}, \b_l$ have been already considered,
we must take into account only the extra generator $\e$ and the
associated extension coefficients. For these, successively we obtain
\be
\begin{array}{lll}
\{\j_{ab},\b_b,\e \}:& \quad \me_{ab}=0 \cr
\{\m_{ab},\b_b,\e \}:& \quad \je_{ab}=0 \cr
\{\j_{k-1 k},\m_{k-1 k},\e \}: & \quad \k_{k} \bee_{k}=0 \cr
\end{array}
\label{cpp}
\ee
so the extension coefficients in (\ref{ECg}) are equal to zero,  and those in
(\ref{ECh}) are basic, to be denoted as
\be
\cextg_k:= \bee_k
\ee
and must satisfy
\be
\k_k \cextg_{k}=0.
\label{cextg.rel}
\ee

Again in this case, it is easy to check that all  remaining Jacobi
equations involving the generator $\e$ are satisfied and  do not lead
to any further relation.


\end{document}